\documentclass[12pt]{article}

\usepackage{graphicx}
\usepackage{dsfont}
\usepackage{amssymb}
\usepackage{amsmath}
\usepackage{mathrsfs}
\usepackage{bm}

\title{
Comment on Self-Stress on a Dielectric Ball and Casimir-Polder Forces}
\author{Ulf Leonhardt\\
Department of Physics of Complex Systems,\\
Weizmann Institute of Science, Rehovot 76100, Israel}

\date{\today}

\begin{document}
\maketitle

\begin{abstract}
In our paper [Ann. Phys. (NY) {\bf 395}, 326 (2018)] we calculate the Casimir stress on a sphere immersed in a homogeneous background, assuming dispersionless dielectrics. Our results appear to challenge the conventional picture of Casimir forces. The paper [arXiv:1909.05721] criticises our approach without offering an alternative. In particular, the paper [arXiv:1909.05721] claims that we have made an unjustified mathematical step. This brief comment clarifies the matter. 
\end{abstract}

The point in question is the following. The force density $\bm{f}$ is calculated as the divergence of Maxwell's stress tensor $\sigma$ evaluated in the vacuum state and renormalized according to standard Lifshitz theory \cite{LL9}. It is clear from symmetry considerations that the force density on the sphere can only point in radial direction. However, in spherical coordinates, there are two contributions to the divergence, one from the radial stress $\sigma_r^r$ and another one from the transversal stress $\sigma_\theta^\theta=\sigma_\phi^\phi$ according to the formula
\begin{equation}
f_r = \frac{1}{r^2}\frac{\mathrm{d} (r^2 \sigma_r^r)}{\mathrm{d} r} - \frac{2}{r}\,\sigma_\theta^\theta \,.
\label{div}
\end{equation}
It is also clear that the force density is concentrated at the interface of the sphere. With $a$ being the radius of the sphere, the total force must be given by
\begin{equation}
4\pi a^2 f_r= F \delta(r-a) \,.
\label{delta}
\end{equation}
In our paper \cite{Yael} we verify by exact calculation that the force density vanishes inside and outside the sphere. In order to determine the force at the interface, we expand $r^2\sigma_r^r$ as Laurant and logarithmic series of the distance $\Delta$ from the interface, one for $r<a$ and one for $r>a$. Note that these two series are different. Otherwise the divergence cannot produce a delta--function singularity --- the delta function is the derivative of the step function, and so the differentiation of a discontinuous function is required in Eq.~(\ref{div}) for generating the surface force of Eq.~(\ref{delta}). From this follows another mathematical fact. While the transversal stress $\sigma_\theta^\theta$ compensates for the contribution of the radial stress $\sigma_r^r$ such that the force vanishes for $r\neq a$, it is solely the radial stress that can contribute to the delta--function singularity at $r=a$, as only $r^2\sigma_r^r$ is differentiated in Eq.~(\ref{div}). 

Paper \cite{Milton} claims that the transversal contribution is ``mysteriously omitted'' and regards this as the main mathematical reason why our paper is allegedly wrong. But there is no mystery, just logic. Admittedly, that logic was not sufficiently clearly expressed in our paper \cite{Yael}. Hence there is a need for this comment to clarify the matter. 

There are also other incorrect claims that are worth correcting. We do not, as claimed \cite{Milton}, omit the divergent terms in the renormalized stress. Instead, we give them a physical meaning as the conventional surface tension. Here microscopic details, in particular the sizes of molecules, will make these forces finite. We calculate what can be calculated with macroscopic electromagnetism: a Casimir contribution to the surface tension. We are aware, and state this in our paper, that the separation between converging and diverging components is not unique for the logarithmic part of the series expansion, but we have found that this does not affect the most surprising consequence of our paper \cite{Yael}: the linear dependence of the force on density in the dilute limit. This puzzling behaviour puts the conventional picture of Casimir forces into question. It has been believed that Casimir forces should be reduced to intermolecular forces in the dilute limit, which would imply that the force always depends quadratically on density. Our paper \cite{Yael} shows that this is not the case. 

So far, this has been theory, and the critics \cite{Milton} do not even offer an alternative,  but there are indications from a related experiment \cite{Matzliah} hinting at a linear density dependence. Here a dilute cloud of ultracold atoms is illuminated with light interacting with the atoms. The light perceives the atoms as a dielectric medium and their cloud as a lens, it is getting focused and hence changes momentum. Meanwhile the atoms pick up the recoil via the optical forces they experience. These forces are fundamentally the same as the Casimir force, except that they are not driven by vacuum fluctuations alone, but by the macroscopic electromagnetic field of the incident light. The measurements \cite{Matzliah} clearly show that the force on the atoms depends linearly on density, not quadratically, as most theorists had expected. 

This experiment \cite{Matzliah} and the agreement of our theory with previous special cases \cite{Yael} give us some confidence. We do not claim to understand the whole story of the Casimir stress in a sphere and we agree that our results are surprising, but they come from physically justified and mathematically correct calculations.


\begin{thebibliography}{99}
\bibitem{LL9}
L. D. Landau and E. M. Lifshitz,
{\it Statistical Physics, Part 2}
(Pergamon, Oxford, 1980).
\bibitem{Yael}
Y. Avni and U. Leonhardt,
Ann. Phys. (NY) {\bf 395}, 326 (2018).
\bibitem{Milton}
K. A. Milton, P. Parashar, I. Brevik, and G. Kennedy,
arXiv:1909.05721.
\bibitem{Matzliah}
N. Matzliah, H. Edri, A. Sinay, R. Ozeri, and N. Davidson,
Phys. Rev. Lett. {\bf 119}, 163201 (2017).
\end{thebibliography}
\end{document}